\documentclass[preprintnumbers,amsmath,amssymb,twocolumn,prl,showpacs]{revtex4-1}

\usepackage{graphicx,comment}% Include figure files

\usepackage{bm}% bold math
\usepackage{hyperref}
\usepackage{mathtools,amsfonts}
\usepackage{algorithm}
\usepackage{setspace}

\def\tl{{\tilde \lambda}}
\def\tbg{{\tilde{ \bf {g}}}}
\def\rf{{\rm f}}

\def\cG{\mathcal{G}}

\def\cK{\mathcal{K}}

\def\cV{\mathcal{V}}

\def\Tr{\mathrm{Tr}}

\newcommand{\ket}[1]{\left| #1\right\rangle}        % ket vector
  
\begin{document}
\title{Quantum eigenvalue estimation via time series analysis}
\author{Rolando D. Somma}
\address{Theoretical Division, Los Alamos National Laboratory, Los Alamos, NM 87545, USA}

\date{\today}

\begin{abstract}
We present an efficient method for estimating the eigenvalues
of a Hamiltonian $H$ from the expectation values of the evolution
operator for various times. For a given quantum state $\rho$,
our method outputs  a list of eigenvalue estimates and approximate  probabilities. Each probability
depends on the support of $\rho$ in those eigenstates of $H$ associated with eigenvalues within an arbitrarily small range. The complexity of our method is polynomial in the inverse of a given precision parameter $\epsilon$, which is the gap between eigenvalue estimates. Unlike the well-known quantum phase estimation algorithm that uses the quantum Fourier transform, our method does not require large ancillary systems, large sequences of controlled operations, or preserving coherence between experiments, and is therefore more attractive for near-term applications.
The output of our method can be used to compute spectral properties of $H$ and other expectation values efficiently, within additive error proportional to $\epsilon$.
\end{abstract}
\maketitle

\section{Introduction}
\label{sec:intro}
\begin{figure}[htb]
    \includegraphics[width=8.5cm]{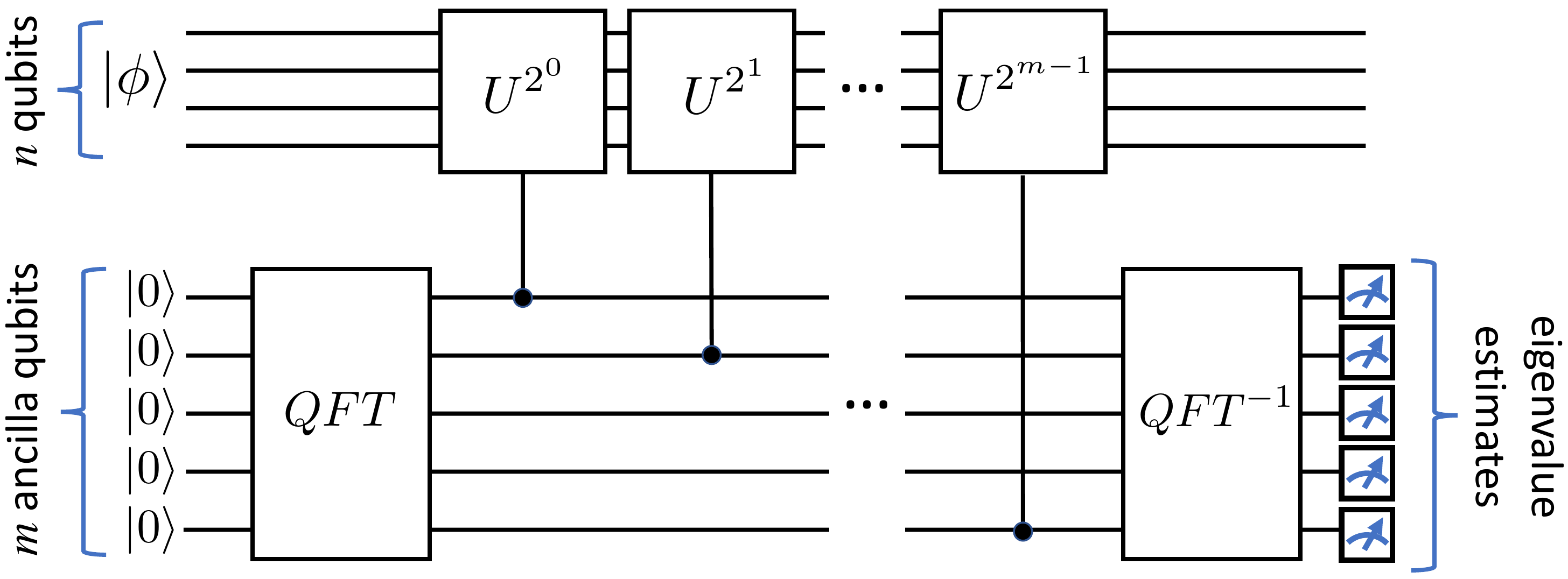}
     \caption{The quantum phase estimation algorithm for estimating the eigenvalues of a unitary operator $U$. $QFT$ is the quantum Fourier transform and the filled circles denote operations controlled on the states $\ket 1$ of corresponding ancilla qubits. The number of these ancilla qubits, $m$, depends on the desired precision in the estimation. Projective measurements on the ancilla qubits provide the eigenvalue estimates.}
    \label{fig:PEA}
\end{figure}

One of the most powerful and widely used quantum algorithms
is quantum eigenvalue or phase estimation (QPE)~\cite{Kit95,CEMM98,NC01} -- see Fig.~\ref{fig:PEA}.
This algorithm allows us to estimate eigenvalues of Hermitian or unitary operators
and plays a key role in quantum computing: it is a subroutine of, for example, Shor's algorithm
for factoring~\cite{Sho97} and algorithms for solving systems of linear equations~\cite{HHL09,Amb12,CKS17}.
QPE is also useful in physics and chemistry~\cite{AL99,ADL+05},
and in quantum metrology~\cite{KOS07,HBB+07}. 

As we enter the era of noisy, intermediate-scale quantum (NISQ) technologies, of significant interest is the development of quantum algorithms that require fewer ancilla qubits, fewer (controlled) quantum gates, fewer measurements, or quantum circuits of shorter depth -- see Refs.~\cite{SHF13,PMS14,KS17,CSSC18} as examples. 
We present another contribution in this regard by
providing an alternative method for performing efficient eigenvalue estimation based on a time series analysis. 
The time series
may be obtained using simple quantum algorithms that require, at most, one ancilla qubit and one single-qubit measurement per run -- see below. This is in sharp contrast
with the former QPE algorithm of Fig.~\ref{fig:PEA}, where the number of ancilla qubits, controlled operations, and many-qubit measurements in a single execution depend on the desired precision and confidence level, if using a high-confidence version of QPE. Additionally,
our method does not use the QFT, thereby avoiding 
the multiple controlled two-qubit gates needed for its implementation.

Furthermore, while  the QPE algorithm of Fig.~\ref{fig:PEA} may still be simulated using one ancilla qubit only following Ref.~\cite{GN96}, that approach requires performing single-qubit measurements at intermediate steps and preserving quantum coherence from one experiment to the next~\cite{KOS07}. This is a strong requirement that will not be needed here. Thus, by eliminating expensive resources, our results may be useful for
NISQ technologies, as long as the
the noise in the hardware is not a limiting factor in achieving the desired precision. To this end, our results may be combined
with those in, e.g., Ref.~\cite{KTC19} to mitigate errors and improve accuracy.

Our approach to eigenvalue estimation
will be particularly useful to physics and chemistry problems,
as many properties can be classically computed after the estimates
are obtained.
In more detail, given an $n$-qubit Hamiltonian $H$, our method
outputs a list of estimates of eigenvalues of $H$, together 
with a list of approximate probabilities. The size of the lists
is determined by a precision parameter $\epsilon>0$. The output can then be used 
to compute various spectral properties of $H$ efficiently,
such as its expected value on a given state $\rho$ or other moments.
The complexity of our method 
is polynomial in $1/\epsilon$ and logarithmic in $1/(1-c)$, where
$c<1$ is the confidence level. This complexity takes into account the total
number of quantum operations,  state preparations, ancillary qubits, one-qubit measurements, classical computations, and total evolution time under $H$ needed to produce the desired output. 

The paper is organized as follows. 
First, we review two previous approaches to quantum eigenvalue
estimation via a time series analysis and emphasize
the advantages of the current approach. We then formulate the quantum
eigenvalue estimation problem (QEEP) in more detail 
and present our solution. 
We show how the output of our method can be used
to compute spectral properties of $H$ within arbitrary
accuracy and confidence level.
We also compare our approach with the one 
that uses the QPE algorithm of Fig.~\ref{fig:PEA} and
present some numerical results that demonstrate
the feasibility of our method. Finally we
provide some concluding remarks.

%%%%%%%%%%%%%%%%%%%%%%%%%%%%%%%%%%%%%%%%%%
%%%%%%%%%%%%%%%%%%%%%%%%%%%%%%%%%%%%%%%%%%
  \subsection{Related work}
\begin{figure}[htb]
    \includegraphics[width=7cm]{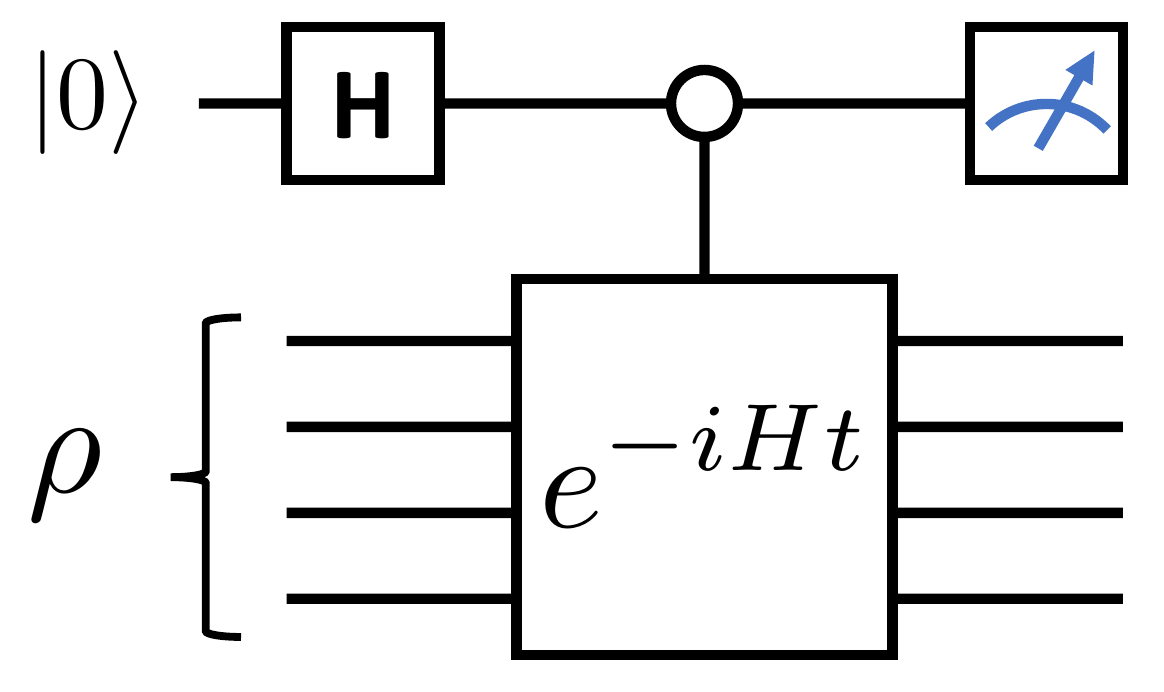}
     \caption{A quantum algorithm to obtain the expectation
     of the evolution operator using only
     one ancilla qubit, initialized in $\ket 0$. $\bf {\rm H}$ denotes
     the Hadamard gate and the circle denotes a controlled operation
     in the state $\ket 1$ of the ancilla. Repeated measurements of the ancilla-qubit Pauli
     operators $\sigma_x$ and $\sigma_y$ result in the expectation
     $\langle \sigma_x + i \sigma_y \rangle=\Tr[\rho.e^{-iHt}]$.}
    \label{fig:1AncillaAlg}
\end{figure}

 The idea of performing eigenvalue estimation
using the time series (TS)
was suggested in Ref.~\cite{SOGKL02}.
The TS is given
 by the expectations of the evolution operator,
i.e. $g(t)=\Tr[\rho.e^{-iHt}]$, for various times $t=t_0 <t_1 <\ldots$.
Here, $\rho$ and $H$ are the state and Hamiltonian of a system of $n$ qubits, respectively. $g(t)$ can be obtained from multiple executions of the one-ancilla quantum algorithm depicted in Fig.~\ref{fig:1AncillaAlg}.
 From the TS, Ref.~\cite{SOGKL02} suggested using the (classical) discrete
 Fourier transform (DFT) to obtain the eigenvalues of $H$ as well as the support of $\rho$
 on the corresponding eigenstates. Intuitively, this approach should work because
 \begin{equation}
    g(t)=\sum_\lambda r_\lambda e^{-i \lambda t} \;,
 \end{equation}
 where the $\lambda$'s are the eigenvalues of $H$ and the $r_\lambda$'s
 are
 the probabilities of $\rho$ being in the corresponding eigenstates.
 Then, the DFT may be used to estimate
 the frequencies (eigenvalues) and components  (probabilities) of the time-dependent ``signal'' $g(t)$.

\begin{figure}[htb]
    \includegraphics[width=8cm]{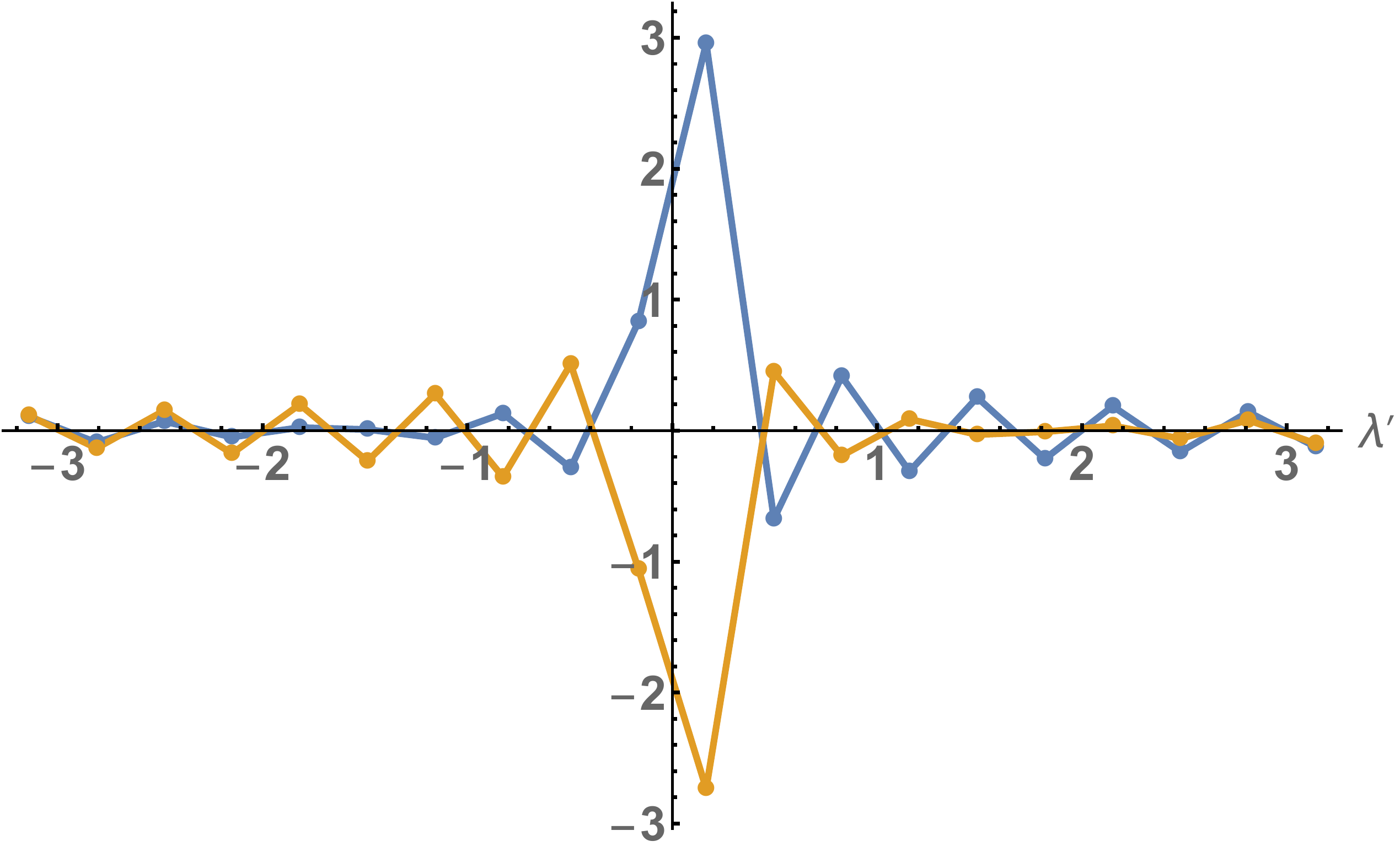}
     \caption{The real (blue dots) and imaginary (orange dots) parts of the
     vector resulting from the action of
     the DFT on ${\bf g}$,
     with $M$ components $g_k=e^{-i\lambda k}$, $k=0,\ldots,M-1$. The dimension is $M=20$ and $\lambda=2\pi/80$ in this case. The scale for $\lambda'$
     is such that $| \lambda'| \le \pi$, corresponding
     to the eigenvalue estimates. The appearance of slowly-decaying Fourier coefficients may result in inaccurate estimates of the eigenvalue if no other processing is performed.}
    \label{fig:DFT_1Frequency}
\end{figure}

 This DFT-based approach may work well
 when the number of distinct eigenvalues
 is small (say, a constant) and when these eigenvalues
 are well separated from each other (say, by constant gaps). However, even in that case,
 obtaining accurate estimates of the $\lambda$'s and $r_\lambda$'s from the Fourier-transformed
 signal may require further post-processing. To explain this, we consider the simplest
 example where only one eigenvalue is present and $g(t)$ is known
 for various times $t_k= k$, $k=0,1,\ldots,M-1$. For simplicity,
 we assume that the eigenvalue satisfies $| \lambda |\le \pi$
 and define $g_k:=g(t_k)$.
 In Fig.~\ref{fig:DFT_1Frequency}, we plot  the result from
 the action of the $M\times M$-dimensional DFT
 on the vector ${\bf g}=(g_0,\ldots,g_{M-1})$. As $\lambda$ is not 
 a multiple of $2\pi/M$ in this example,  all the values in the plot are nonzero and, while it is evident that
 $\lambda \approx 0$ from the plot, estimating
 the actual value of $\lambda$ requires additional calculations~\cite{SOGKL02}. The situation is far more complex
 when the number of distinct eigenvalues is large, such as when this number scales
 exponentially in $n$,
 and when the $g_k$'s are only approximately known. Simple attempts to solve this general case will be inefficient, likely resulting in undesirably large complexities. Remarkably, the current approach to eigenvalue estimation will overcome the disadvantages of the DFT-based approach.

In Ref.~\cite{BTT19}, the authors provide a method for eigenvalue
estimation, which uses the matrix pencil (MP) method~\cite{SP95}, and is also based on a time series analysis. The focus is on computing
the lowest energy of the Hamiltonian or computing many eigenvalues,
under the assumption that the number of distinct $\lambda$'s is not too large.
The claims of Ref.~\cite{BTT19} are mostly based on numerical simulations.
For the problem of computing the lowest eigenvalue or energy, $\lambda_m$, the authors observe
that the complexity of the MP approach is polynomial in $1/r_{\lambda_m}$ and the inverse of the gap between $\lambda_m$
and the nearest eigenvalue. For the problem of computing multiple
eigenvalues, the complexity is also polynomial in the number of distinct
eigenvalues. Naturally, these complexities will be extremely large
for typical cases where the gap is exponentially small in $n$. See Refs.~\cite{SRM17,BT19}
for related uses of the MP method in quantum computing.

Our method aims at solving a different, but related, problem.
For the QEEP, we split the range of eigenvalues into bins of a given size,  corresponding to the various eigenvalue estimates. The resulting probabilities depend
on the support of $\rho$ in the eigenstates associated with specific bins. (We note that this is similar to the eigenvalue estimation problem that can be solved by using the QPE algorithm of Fig.~\ref{fig:PEA}.)
In a way, our goal is less ambitious than
computing single eigenvalues, reason why our 
method is efficient even when the number of distinct eigenvalues
is exponential in $n$. Our method, however, can also be extended to solve the
problem of Ref.~\cite{BTT19}, and we provide analytical
bounds that are particularly useful when the vector ${\bf g}$ is not exactly known. 

We note that the MP approach, while lacking analytical bounds of its convergence in the noisy case, can in principle 
be used to estimate multiple eigenvalues as well. A natural question is  how that method compares
with the current TS approach. To this end, we will present some numerical simulations
that show that our TS approach outperforms the MP approach in computations of certain expectations of $H$.

%%%%%%%%%%%%%%%%%%%%%%%%%%%%%%%%%%%%%%%%%%%%%%%%%%%%%%%%
%%%%%%%%%%%%%%%%%%%%%%%%%%%%%%%%%%%%%%%%%%%%%%%%%%%%%%%%
\section{Quantum eigenvalue estimation problem (QEEP)}
We present the QEEP in more detail. As before, we assume that $H$ and $\rho$
represent the (dimensionless) Hamiltonian and quantum state of a system of $n$ qubits, respectively. 
With no loss of generality, we let $\|H\|\le 1/2$ so that its eigenvalues satisfy $-1/2 \le \lambda \le 1/2$.
Given a precision parameter $\epsilon >0$, which can be associated with (half of) a bin size, the eigenvalue
estimates are $\tilde \lambda_j :=-1/2 + j \epsilon$, $0 \le j \le M-1$, where $M=1+1/\epsilon$~\cite{note1}.
In addition, the goal is to compute a vector ${\bf q}=(q_0,q_1,\ldots,q_{M-1}) \in \mathbb R^M$ satisfying
\begin{align}
\label{eq:QEEPsol}
\| {\bf q} - {\bf p} \|_1 \le \epsilon \;,
\end{align}
with probability at least $c>0$. Here, $\|.\|_1$
denotes the $L^1$-norm.
The probability vector ${\bf p}=(p_0,p_1,\ldots,p_{M-1})$
is such that
\begin{equation}
\label{eq:p_jdef}
    p_j = \sum_{\lambda \in \cV_j } f_j(\lambda) r_\lambda \;,
\end{equation}
where $\cV_j=[-1/2+(j-1) \epsilon,-1/2+(j+1)\epsilon]$ refers to a particular bin and
$r_\lambda$ is the probability of $\rho$ being in the eigenstate $\ket{\psi_\lambda}$.
That is, $p_j$ only depends on the support of $\rho$
in the eigenspace associated with the eigenvalues lying
in the $j$-th bin, i.e.,  the eigenvalues that are $\epsilon$-close to the estimate $\tl_j$. 
For   $0 \le j \le M-2$,
the functions $f_j(\lambda)$ are non-negative and satisfy
\begin{equation}
\label{eq:fnormalization}
    f_j(\lambda)+f_{j+1}(\lambda)=1 \;, 
\end{equation}
for all $\lambda \in \cV_j \cap \cV_{j+1}$. 

In the previous definition of the QEEP, the bins $\cV_j$  are of size $2\epsilon$
and adjacent bins overlap in a region of size $\epsilon$.
The condition of Eq.~\eqref{eq:fnormalization}
is then necessary to avoid double counting and to satisfy the rule $\sum_{j=0}^{M-1} p_j=1$.
The reason why we consider overlapping bins is to avoid undesired complexity overheads and this will become
clear in the next section. For example, if the bins did not overlap, complications could arise from those 
eigenvalues that are very close to the boundary of a bin. 
Additionally, overlapping bins will allow us to exploit a property of smooth functions $f_j(\lambda)$, 
namely the rapid decay of their Fourier coefficients--see next.

Our definition of the QEEP is also motivated by the QPE algorithm of Fig.~\ref{fig:PEA}
and its high-confidence variant~\cite{KOS07} . 
In that case, when the $n$-qubit input state $\ket \phi$ is an eigenstate of $H$ of eigenvalue $\lambda$, 
and when the desired confidence level approaches 1, the QPE algorithm outputs one of the two closest 
eigenvalue estimates, $\tl_j$ or $\tl_{j+1}$, with almost probability one.
(Note that $\tl_j \le \lambda \le \tl_{j+1}$.)
The probability of any of these estimates is also determined by functions $f_j(\lambda)$, 
which can be obtained by analyzing the action of the QPE algorithm on the eigenstate.

%%%%%%%%%%%%%%%%%%%%%%%%%%%%%%%%%%%%%%%%%%%%%%%%%%%%%%%%
%%%%%%%%%%%%%%%%%%%%%%%%%%%%%%%%%%%%%%%%%%%%%%%%%%%%%%%%
\section{Solution to the QEEP from the time series}
\label{sec:solution}
While the QEEP may be solved using a variety of methods~\cite{CEMM98,SHF13,BTT19},
in this paper we are interested in an efficient method that finds a solution
using a TS approach. To this end, we assume
that $\tbg$ is an estimate of $\bf g$; that is, an $N$-dimensional (random) vector that satisfies
\begin{equation}
\label{eq:L1bound0}
    \| \tbg - {\bf g}\|_1 \le \epsilon \;,
\end{equation}
with probability at least $c$. 
The components of $\bf g$ are $g_k=\Tr[\rho.e^{-iHk}]$, $k=0,1,\ldots,N-1$, and 
we assume $\tilde g_0=1$.
Here, $N$ depends on the precision parameter $\epsilon$ and will be determined below.
The goal is to obtain the vector ${\bf q}$ in Eq.~\ref{eq:QEEPsol}, where each $q_j$ will result
from a particular linear combination of components of $\tbg$.

For $x \in \mathbb R$ and $0 \le j \le M-1$, we define the indicator functions
\begin{align}
   { 1}_j(x):= 1 , \; {\rm if} \; \tl_j -\frac \epsilon 2 \le x < \tl_j + \frac \epsilon 2 \; , \\
    { 1}_j(x):= 0  , \; {\rm otherwise} \; .
\end{align}
Note that a possible set of functions $f_j(\lambda)$ that satisfy Eq.~\ref{eq:fnormalization} can be obtained
from these indicator functions. Similarly, we can define the operators
$\hat { 1}_j(H)$ such that $\hat { 1}_j(H)\ket{\psi_\lambda}={ 1}_j(\lambda)\ket{\psi_\lambda}$,
where $\ket{\psi_\lambda}$ is the eigenstate of $H$ of eigenvalue $\lambda$.
That is, $\hat 1_j(H)$ acts as the projector onto the eigenspace
spanned by eigenstates $\ket{\psi_\lambda}$ with eigenvalues
in the corresponding region. Then, an exact solution to the QEEP could be obtained by computing the expectations
$\Tr[\rho.\hat 1_j(H)]$, for all $0 \le j \le M-1$.
A Fourier approach would allow us to decompose each $\hat 1_j(H)$  as a combination of operators like $e^{-iHt_k}$,
and the previous expectations could be obtained, in principle, 
from the time series.

However, as the Fourier coefficients of the indicator function
decay slowly (as $\sim 1/|k|$), the previous approach would require significant
resources for solving the QEEP.
For example, it would require computing the expectation values
of $e^{-iHk}$ for undesirably large values of $k$ that scale with $1/\epsilon^2$ (or worse).
To avoid this problem, we can consider another set of functions that
are smooth in the corresponding intervals while still satisfying Eq.~\ref{eq:fnormalization}. 
The Fourier coefficients of smooth functions decay rapidly.
One such a set can be obtained from
the so-called bump function as follows. Let
$h(x):=a.\exp(-1/(1-x^2))$ if $|x|<1$,
where $a \approx 2.25$ is for normalization purposes, and $h(x)=0$ if $|x| \ge 1$. The function $h(x)$ is smooth for $x \in \mathbb R$ and we
also define $ h_\epsilon(x):= \frac 2 \epsilon h(2x/\epsilon)$,
which is nonzero only when $-\frac \epsilon 2 \le x \le \frac \epsilon 2$.
Last, we define the functions
\begin{equation}
\label{eq:fsmooth}
f_j(x):= \int_{-\infty}^\infty dx' \; h_\epsilon(x'-x) \; { 1}_j(x') \;.
\end{equation}

In Fig.~\ref{fig:Functions} we plot the functions
$f_j(x)$. Their relevant properties are analyzed in Appendix A.
The support of $f_j(x)$ is $\cV_j$. Additionally, $f_j(x) \ge 0$,
$f_j(x)+f_{j+1}(x)=1$ for all $x \in \cV_j \cap \cV_{j+1}$,
and the property of Eq.~\ref{eq:fnormalization} is satisfied,
after replacing $x$ by the eigenvalue $\lambda$.

 \begin{figure}[htb]
    \includegraphics[width=8.5cm]{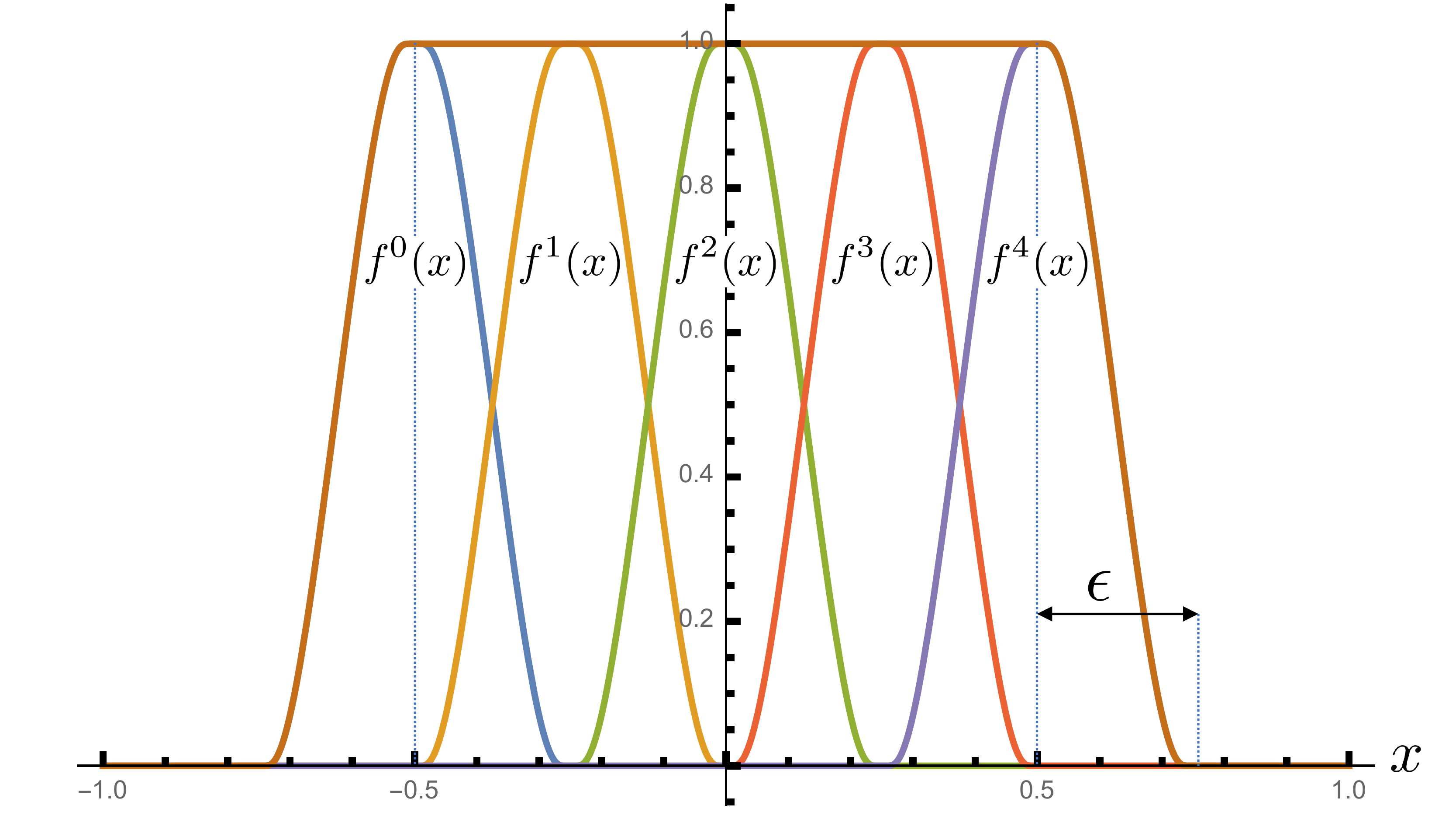}
     \caption{The functions $f_j(x)$ for $0 \le j \le 4$ ($M=5$) and their sum (brown line). Here, $\epsilon=1/4$ and the eigenvalue estimates are $\tl_j=-1/2+j/4$.
     Each $f_j(x)$ has compact support in $\cV_j=[-3/4+j/4,-1/4+j/4]$. As the sum of these functions is 1 in the relevant region $-1/2 \le x \le 1/2$, the $f_j(x)$'s
     satisfy Eq.~\ref{eq:fnormalization} after replacing $x \rightarrow \lambda$.}
    \label{fig:Functions}
\end{figure}

As before, we can define the operators $\hat f_j(H)$ such that
$\hat f_j(H) \ket{\psi_\lambda}=f_j(\lambda)\ket{\psi_\lambda}$.
Then, an exact solution to the QEEP follows from the $M$ expectation
values $p_j = \Tr[\rho.\hat f_j(H)]$.
As the functions
$f_j(x)$ are smooth, their Fourier coefficients must decay superpolynomially
fast~\cite{MD73}. After approximating $\hat f_j(H)$
by linear combinations of $e^{-iHk}$,
the $p_j$'s could be well approximated from the time series $\bf g$,
where the largest evolution time will be reasonably bounded -- see below. 

For our analysis, we find it convenient to actually use the periodic functions
$\rf_j (x):=\sum_m f_j(x+m2\pi)$ and the corresponding
operators $\hat \rf_j (H)$. As the eigenvalues of $H$
satisfy $|\lambda| \le 1/2$, we also obtain
$p_j = \Tr[\rho . \hat \rf_j (H)]$.
The Fourier series implies
\begin{equation}
\hat \rf_j(H) = \frac 1 {\sqrt{2\pi}} \sum_{k=-\infty}^\infty   F_j(k) e^{i H k}  \; ,
\end{equation}
where $F_j(k)$ is the Fourier transform of $f_j(x)$. In Appendix~A
we show that $|F_j(k)|$ decays rapidly
in the limit of large $k$. As we seek to avoid evolving with $H$ for large times,
we can then approximate $\hat \rf_j(H)$ by dropping the terms in the sum
for $|k| \gg 1$.  To this end, we define the approximate operators
\begin{equation}
\hat \rf '_j(H) := \frac 1 {\sqrt{2\pi}} \sum_{|k|<N}   F_j(k) e^{i H k} \;,
\end{equation}
for some given $N \gg 1$, and also define a vector ${\bf p}'$
with components $p'_j:=\Tr[\rho. \hat \rf'_j (H)]$.
 In Appendix~A we show that there exists
$N=O(\log^2(1/\epsilon)/\epsilon)$
such that
\begin{equation}
\label{eq:L1bound1}
    \|{\bf p'} - {\bf p}\|_1\le \epsilon/2 \;.
\end{equation}

The components $p'_j$ can be obtained from the time series as
\begin{equation}
\label{eq:p'}
    p'_j=\frac 1 {\sqrt{2\pi}} \sum_{|k|< N} F_j(k) g_k^* \;,
\end{equation}
where we used the property $g_{-k}=g_{k}^*$.
Since we only have an estimate of $\bf g$,
our solution to the QEEP is determined by
a vector $\bf q$ of components
\begin{align}
\nonumber
    q_j &:= \frac 1 {\sqrt{2\pi}} \sum_{|k|< N} F_j(k) \tilde g_k^*  \\
    \label{eq:q}
    & = \frac {\epsilon}{2 \pi} + \sqrt{\frac 2 \pi} \Re \left( \sum_{k=1}^{N-1}F_j(k) \tilde g_k^* \right)\;,
\end{align}
where we defined $\tilde g_{-k}=\tilde g_{k}^*$, for $k \ge 0$,
and used the properties $F_j(-k)=(F_j(k))^*$ and $F_j(0)=\epsilon/\sqrt{2\pi}$.

From Eq.~\ref{eq:L1bound0}, our solution to the QEEP satisfies
\begin{align}
    \|{ \bf q} - {\bf p'} \|_1 &\le \frac 1 {\sqrt{2\pi}} \sum_{j=0}^{M-1} \sum_{|k|< N} |F_j(k)| |\tilde g_k^* - g_k^*| \\
    \label{eq:L1bound2}
    & \le  \epsilon /2 \;,
\end{align}
with confidence level bounded by $c$.
The last inequality in Eq.~\ref{eq:L1bound2} was obtained using
$M=1+1/\epsilon \le 2/\epsilon$,
and $|F_j(k)| \le \epsilon/(2\pi)$ -- see Appendix~A.
Equations~\ref{eq:L1bound1} and~\ref{eq:L1bound2} imply the desired condition
of Eq.~\ref{eq:QEEPsol}.

%%%%%%%%%%%%%%%%%%%%%%%%%%%%%%%%%%%%%%%%%%%%%%%%%%%%%%%%
%%%%%%%%%%%%%%%%%%%%%%%%%%%%%%%%%%%%%%%%%%%%%%%%%%%%%%%%
\subsection{Computation of $\tbg$}
To obtain the desired vector $\bf q$, 
we need to estimate the expectations of $e^{-iHk}$
in the state $\rho$, for $1 \le k <N$. These estimates
are required to satisfy
Eq.~\ref{eq:L1bound0}. For simplicity,
we assume that each such estimation is done within the
same absolute precision $\epsilon'=\epsilon/N$ and confidence level $c' \ge 1 - (1-c)/N$, but this assumption
may be relaxed. Under this assumption, the overall confidence level is $c'^N \ge c$.
For our choice of $N$, this implies $\epsilon'=\tilde O(\epsilon^2)$,
where the $\tilde O$ notation hides factors that are logarithmic in
$1/\epsilon$. Each $\tilde g_k$ can be obtained by repeated use of
the algorithm of Fig.~\ref{fig:1AncillaAlg}.

%%%%%%%%%%%%%%%%%%%%%%%%%%%%%%%%%%%%%%%%%%%%%%%%%%%%%%%%
%%%%%%%%%%%%%%%%%%%%%%%%%%%%%%%%%%%%%%%%%%%%%%%%%%%%%%%%
\subsection{Complexity}
The complexity of our procedure is determined by the quantum and classical resources required to obtain ${\bf q}$. These resources include the number of elementary quantum operations (including those
needed to simulate the evolution under $H$),  number of qubits, number of state preparations,
number of measurements, and other computation steps that we now analyze. 

First, we determine the total number of uses $R$ of the
method of  Fig.~\ref{fig:1AncillaAlg} to obtain ${\bf \tilde g}$.
Hoeffding's inequality
implies that, to achieve absolute precision $\epsilon'$ and confidence level $c'$ in the estimation of each $g_k$,
$R=O(N |\log(1-c')|/(\epsilon')^2)$  suffices.  
Equivalently, $R=\tilde O(|\log(1-c)|/\epsilon^5)$.
We note that $R$ is also the number of preparations or copies of $\rho$ needed
 as well as the number of projective
one-ancilla measurements in our approach.

The algorithm of Fig.~\ref{fig:1AncillaAlg} uses
the (controlled) unitary $e^{-iHk}$.
Implementing $e^{-iHk}$ on a quantum computer can be done
using two-qubit gates via a variety of quantum
simulation methods~\cite{BAC07,WBH+10,BCC+14,BCC+15,Som16,LC16,LC17}.
Perhaps of most interest for this work are implementations based on the so-called Trotter-Suzuki formula~\cite{Suz90,Suz91}.
These implementations use an approximation of $e^{-iHk}$ by a sequence of evolutions with simpler Hamiltonians and, in contrast
with more recent techniques, they do not 
require additional qubits. To achieve the desired precision, it then suffices to approximate $e^{-iHk}$ within additive precision $\epsilon''=O(\epsilon')$.
This results in a gate complexity (number of two-qubit gates) that is given by $\cG(k,\epsilon')$.
Under some standard assumptions on $H$, the gate complexity is almost linear in $|k|$ and 
the dependence on $1/\epsilon'$ is small (i.e., $\ll 1/\epsilon$). 
Nevertheless, detailed analyses of  gate complexities for
the specific implementations of $e^{-iHk}$ are outside the scope of this
paper. 
As $|k|<N$, the overall gate complexity
of our approach is then bounded by $\cG(N,\epsilon'').R$.  This is roughly
$\tilde O(|\log(1-c)|/\epsilon^6)$, if high-order approximations to $e^{-iHk}$ are used.

To compute $\bf q$,
we need to know the Fourier components $F_j(k)$.
Here, we do not consider the classical cost of obtaining
the $F_j(k)$'s and assume that we can access them via a given lookup table~\cite{note2}. However, computing all the $q_j$'s
from the $\tilde g_k$ has additional classical complexity $O(M.N)=\tilde O(1/\epsilon^2)$, resulting from standard matrix multiplication algorithms.

%%%%%%%%%%%%%%%%%%%%%%%%%%%%%%%%%%%%%%%%%%%%%%%%%%%%%%%%
%%%%%%%%%%%%%%%%%%%%%%%%%%%%%%%%%%%%%%%%%%%%%%%%%%%%%%%%
\section{Spectral properties}
One of the prime uses of our quantum eigenvalue estimation
approach is for the computation of spectral properties. Once the 
vector ${\bf q}$ is obtained, expectation values
can be computed classically as follows. We let
$T(x)$ be a function and $\hat T(H)$ the associated 
operator
after replacing $x$ by $H$. A common task is 
to compute the expectation $\tau = \Tr[\rho . \hat T(H)]$.
For example, when $\hat T(H)=H^s$, $s>0$, the expectation can
provide 
some high order moment. This is in contrast to 
variational quantum algorithms~\cite{PMS14} that
provide information about the expectation of $H$ only.

We will obtain an estimate of $\tau$
 as
\begin{equation}
  \tilde \tau:=  \sum_{j=0}^{M-1} q_j T(\tl_j) \;.
\end{equation}
Clearly, the approximation error will depend on 
the precision parameter $\epsilon$ and decreases
as $\epsilon \rightarrow 0$. In Appendix~B,
we show
\begin{equation}
\label{eq:expectederror}
    \left |\tilde \tau -   \tau \right| \le \epsilon \left( \max_\lambda |T(\lambda)| + \max_\lambda |T'(\lambda)| \right) \;,
\end{equation}
where $T'(\lambda)=\partial_\lambda T(\lambda)$~\cite{note3}.
As $\bf q$ approximates $\bf p$ with confidence level at least $c$,
Eq.~\ref{eq:expectederror} is satisfied also with the same confidence level. If one seeks an estimate of $\tau$ with absolute precision
$\delta>0$, then Eq.~\ref{eq:expectederror} can be used to set a corresponding bound on $\epsilon$.

The QEEP assumes that $H$ is dimensionless. When considering physical systems, where the Hamiltonian $H_{phys}$
has dimensions of energy, we can use $H= H_{phys}/(2\|H_{phys}\|)$,
which satisfies the assumptions of the QEEP. 
We can then compute spectral properties of $H_{phys}$ by computing those of $H$
and including the relevant energy factors in the calculation. 
For example, we can estimate the expectation of $(H_{phys})^s$ from that of $H^s$,
multiplying the result by $(2\|H_{phys}\|)^s$. This operation will also affect the overall additive error 
and should be considered at the time of choosing $\epsilon$.

%%%%%%%%%%%%%%%%%%%%%%%%%%%%%%%%%%%%%%%%%%%%%%%%%%%%%%%%
%%%%%%%%%%%%%%%%%%%%%%%%%%%%%%%%%%%%%%%%%%%%%%%%%%%%%%%%
\section{Comparison with standard quantum phase estimation}
The standard QPE approach based on the algorithm of Fig.~\ref{fig:PEA}
 also provides an estimate of a different probability vector $\bf p$,
which can be used to estimate eigenvalues and other spectral properties. To simplify the analysis,
we consider a sufficiently high-confidence version of QPE~\cite{KOS07}
and disregard complexity overheads that depend on the confidence level.
Each execution of the algorithm will then output an estimate $\tilde \lambda_{j}$ with some  probability $p_{j}$. The estimate of $p_j$ is
$q_{j}$ and can be
  obtained via frequency counts. 
To satisfy Eq.~\ref{eq:QEEPsol}, it then suffices to satisfy $|q_{j}-p_{j}| \le \epsilon/M \le \epsilon^2$ for all $j$. This
would require running the algorithm $\tilde O(1/\epsilon^4)$ times. As
$\tilde \lambda_{j}$ can be represented in binary form using $O(\log(1/\epsilon))$ bits, the number of one-ancilla measurements is
$\tilde O(1/\epsilon^4)$. Additionally, each execution of the algorithm also requires simulating $e^{-iHt}$ for time $t=\tilde O(1/\epsilon)$
and with additive precision $\epsilon''= O(\epsilon')$, which can be done with gate complexity $\cG(t,\epsilon'')$.
The resulting gate complexity of this approach is then $\tilde O(\cG(t,\epsilon'')/\epsilon^4)$.
This represents a slight improvement over the gate complexity of our current approach; both complexities differ by $\tilde O(1/\epsilon)$  factor.
The basic reason for this improvement
is that, with standard QPE, multiple probabilities can be estimated from the same measurement outcomes. 
In the current approach, each $g_k$ has to be estimated individually instead.

%%%%%%%%%%%%%%%%%%%%%%%%%%%%%%%%%%%%%%%%%%%%%%%%%%%%%%%%
%%%%%%%%%%%%%%%%%%%%%%%%%%%%%%%%%%%%%%%%%%%%%%%%%%%%%%%%
\section{Numerical analyses}
\label{sec:NumSim}
We investigate the performance of our TS method and compare it with that of the MP technique
in Ref.~\cite{BTT19} via numerical simulations. To this end, we assume that $\rho$ has nonzero support on $D \le 2^n$ eigenstates, so that
\begin{equation}
g_k = \sum_{d=0}^{D-1} r_{\lambda_d} e^{-i \lambda_d k} \;.
\end{equation} 
Here,  $r_{\lambda_d}$ and $\lambda_d$ are the nonzero probabilities and the eigenvalues, respectively.
These will be randomly sampled while still satisfying the constraints $-1/2\le \lambda_d \le 1/2$ and $\sum_{d=0}^{D-1} r_{\lambda_d}=1$. 
We then add noise to our signal and assume that $\tilde g_k = g_k + \eta_k$, where the $\eta_k$'s are independent and identically distributed
random variables. In particular, we choose $\eta_k$ to be a complex number with a random phase in $[0,2\pi]$, and random magnitude in $[0,\epsilon']$,
for some given $\epsilon' \ge 0$.  Other noise models can also be considered.

We aim at obtaining an estimate
of the expectation $\tau=\Tr[\rho.H^s]$ within a given accuracy $\epsilon>0$, and $s \ge 0$. 
This $\epsilon$ determines $M$ and $N$, the number of bins and the dimension of ${\bf g}$, respectively. For our method, we first compute
the $M$-dimensional vector ${\bf q}$ of components $q_j$ as given by Eq.~\ref{eq:q}. The values
of $F_j(k)$ can be simply obtained by performing the corresponding integrals numerically. Once ${\bf q}$ is computed, 
our estimate of $\tau$ is
\begin{equation}
\label{eq:ttTS}
\tilde \tau _{\rm TS} = \sum_{j=0}^{M-1} q_j  (\tilde \lambda_j)^s \;,
\end{equation}
where $\tilde \lambda_j = -0.5+j \epsilon$.  

The MP method~\cite{SP95,BTT19} can also be used to obtain $L \le N-1$ eigenvalue 
estimates as well as estimates of $r_\lambda$. 
 That method proceeds by first constructing two
Hankel matrices, ${\bf H}^{0,1}$, of dimension $L \times (2N-L-1)$. The entries of these matrices are
\begin{equation}
| {\bf H}^a |_{ll'} = \tilde g_{l+l'+a-N+1} \;,
\end{equation}
where $a \in \{0,1\}$, $0 \le l \le L-1$, and $0 \le l' \le (2N-L-2)$. Note that, for $k<0$, we use $\tilde g_k = (\tilde g_{-k})^*$.
Then, we construct an $L \times L$ matrix $\cK$ via minimization of $\| \cK {\bf H}^0 - {\bf H}^1 \|$, which can be carried 
by a least squares procedure. The $L$ eigenphases of $\cK$ are then the $L$ eigenvalue estimates
of this approach, which we also write as $\tilde \lambda_0, \ldots, \tilde \lambda_{L-1}$. Last, we construct an $L \times L$
matrix $B$ with entries $|B|_{ll'} = e^{-i  \tilde \lambda_{l'} l}$, where $0 \le l,l' \le L-1$. The estimated probabilities
are obtained by minimizing $\| B {\bf q}' -\tilde {\bf g_0} \|$, where ${\bf q}'=(q'_0,\ldots,q'_{L-1})$ is the solution  and
$\tilde {\bf g_0}=(\tilde g_0, \tilde g_1,\ldots, \tilde g_{L-1})$ is obtained from $\tilde {\bf g}$.
The minimization can also be carried via a least squares procedure. After these estimates are obtained,
we approximate $\tau$ as
\begin{equation}
\label{eq:ttMP}
\tilde \tau_{\rm MP} = \sum_{l=0}^{L-1} q'_l (\tilde \lambda_l)^s \;.
\end{equation}
We will set $L=N-1$ in our simulations.  
A detailed analysis of the MP approach and its applications to signal processing (i.e., frequency or eigenvalue estimation) can be found in Ref.~\cite{SP95}.

\begin{figure}[htb]
    \includegraphics[width=8.5cm]{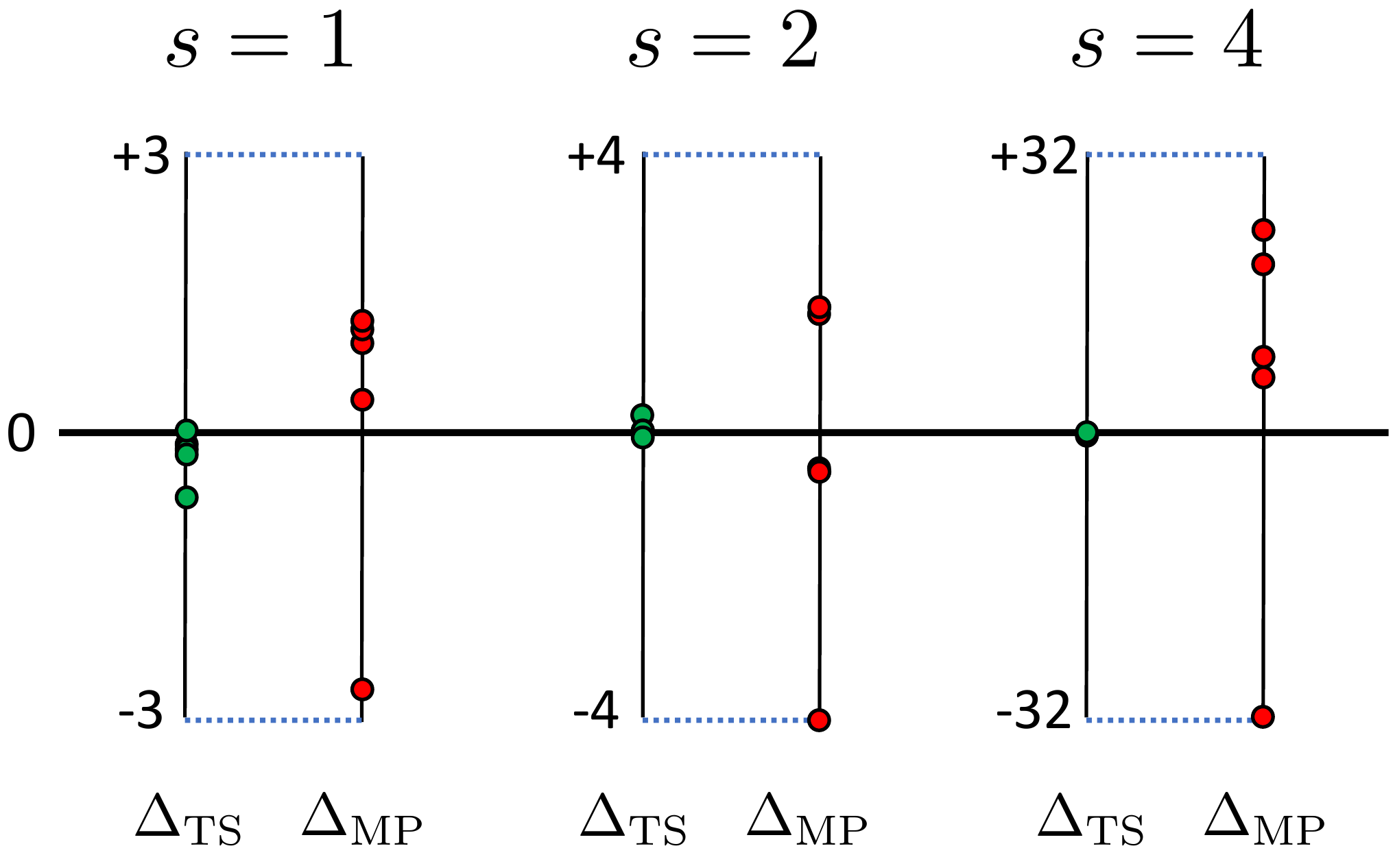}
     \caption{Numerical results for the estimation of the expectation values $\tau=\Tr[\rho.H^s]$, for $s=1,2,4$, from 5 runs. 
     Here, $\Delta_{\rm TS}= (\tau - \tilde \tau_{\rm TS})/\epsilon$ and $\Delta_{\rm MP}= (\tau - \tilde \tau_{\rm MP})/\epsilon$. 
     The chosen parameters are $\epsilon=\epsilon'=0.005$, $M=201$, $N=566$, and $D=5$ distinct eigenvalues. The noisy signal $\tilde {\bf g}$
     was simulated according to the noise model described in the text.}
    \label{fig:NumSim}
\end{figure}

For a fair comparison, both methods use the same vector $\tilde {\bf  g}$
in our simulations.
Numerical simulations show that the MP approach outperforms our current TS approach
in the case where sampling noise is omitted and $\epsilon'=0$. In that case, the MP
approach outputs the exact estimates while the TS approach provides estimates
within the desired accuracy $\epsilon$. Nevertheless, in a realistic scenario
when noise is considered and $\epsilon'>0$,
we observe that the current approach provides significantly better estimates to the expectations $\tau$
than those obtained via the MP method. This is illustrated in Fig.~\ref{fig:NumSim}, where we plot
the errors of the estimates following both, the TS and MP approaches, and considering the
cases where $s=1,2,4$. The  data used for Fig.~\ref{fig:NumSim} is provided in Appendix~C.

\begin{figure}[htb]
    \includegraphics[width=8.5cm]{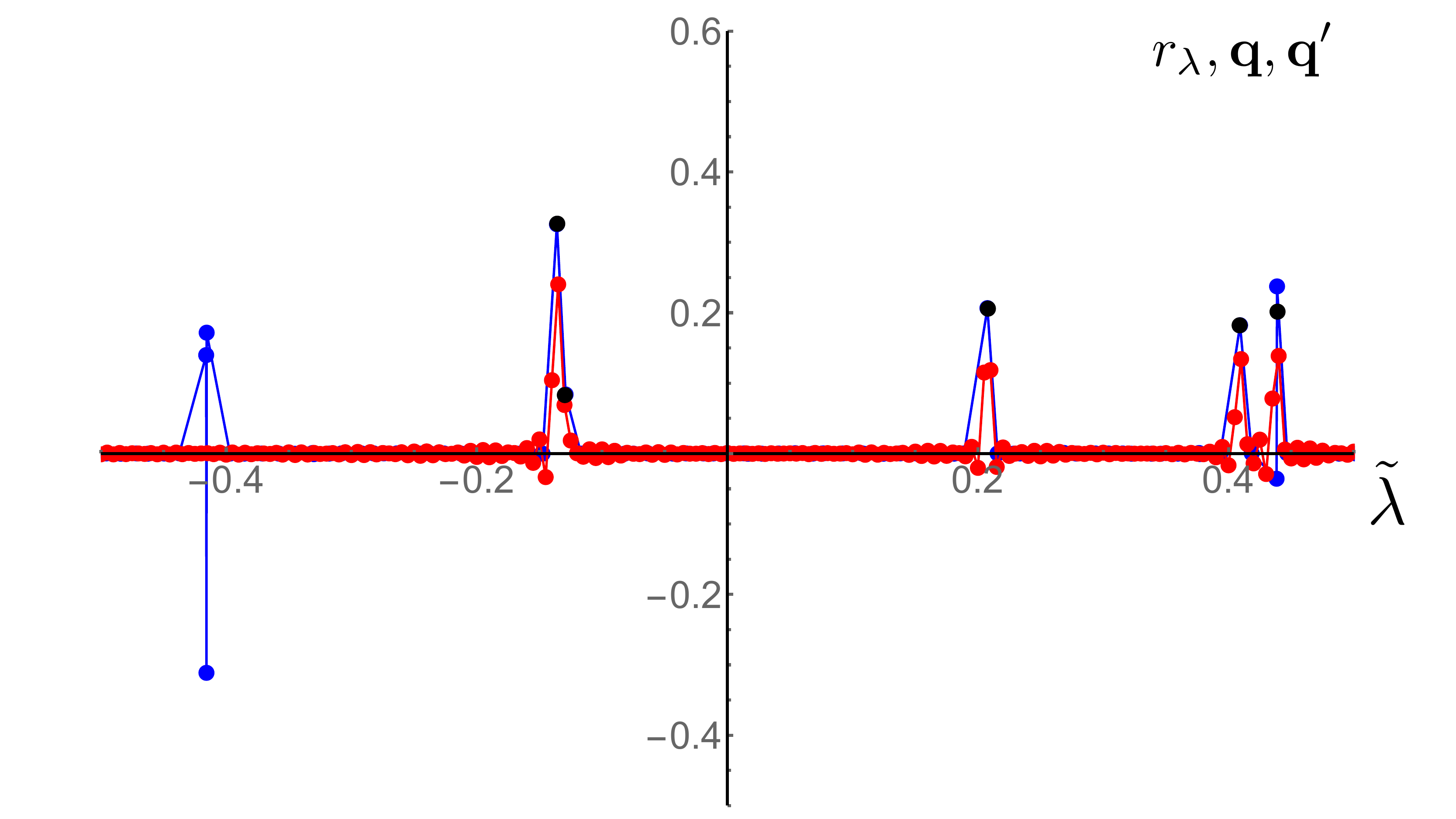}
     \caption{Numerical results for the estimation of eigenvalues in a case where the state $\rho$ is only
     supported on $D=5$ eigenstates. The horizontal axis represents the eigenvalue estimates and the vertical axis represents the estimated probabilities.
     We use the same parameters as for the simulations shown in Fig.~\ref{fig:NumSim},
     where $\epsilon=\epsilon'=0.005$, $M=201$, and $N=566$. The noisy signal $\tilde {\bf g}$
     was simulated according to the noise model described in the text.
     The black dots denote the exact distribution of eigenvalues, $r_\lambda$. The red dots are the results obtained using our TS approach, the vector ${\bf q}$,
     and the blue dots are the results obtained using the MP approach, the vector ${\bf q}'$. The latter provided some spurious results near $\tilde \lambda=-0.4$
     in this case, where $r_\lambda=0$.}
    \label{fig:NumSimQEEP}
\end{figure}

In fact, errors for the MP approach are observed to increase with $s$.
 A possible reason is because that approach
may output eigenvalue estimates outside the range $[-0.5,0.5]$ with non-negligible amplitude.
This contrasts the current TS approach where all eigenvalue estimates are in $[-0.5,0.5]$. However, even if
the estimated eigenvalues outside this range are discarded for the MP approach, the results
do not improve significantly. We plot the computed distribution of eigenvalues in Fig.~\ref{fig:NumSimQEEP}
for the purpose of comparison.
While further post-processing may be able to improve the results of the MP approach, 
that analysis is outside the scope of this paper. 

If the noise in the signal is only due to sampling noise, 
Hoeffding's inequality implies that, to obtain the $N$ components within precision $\epsilon'$ and overall confidence $c$,
$R=(2N/\epsilon'^2) \log(2N/(1-c))$ samples (measurements with $\pm 1$ outcomes)  suffice.   Considering the values of
$N$ and $\epsilon'$ used to obtain the results in Figs.~\ref{fig:NumSim} and~\ref{fig:NumSimQEEP}, and setting
$c=0.99$, we note that $R \approx 5 \ 10^8$ in this case. This is within reach using current quantum technologies~\cite{KMT+17}.
Nevertheless, as $R$ scales poorly with $\epsilon'$, precise-measurement methods such as the one in
Ref.~\cite{TMCM17} or an approach like the one in Ref.~\cite{ACSC20} 
will be important to achieve the desired precision with fewer samples.

%%%%%%%%%%%%%%%%%%%%%%%%%%%%%%%%%%%%%%%%%%%%%%%%%%%%%%%%
%%%%%%%%%%%%%%%%%%%%%%%%%%%%%%%%%%%%%%%%%%%%%%%%%%%%%%%%
\section{Discussions}
We presented a method for quantum eigenvalue estimation
 that avoids or uses significantly less expensive resources than other approaches.
These expensive resources are controlled quantum operations,  ancillary qubits, or the need of preserving coherence from one experiment to the next. 
In addition, the complexity of our method is only slightly
higher than the complexity of the standard QPE approach based on Fig.~\ref{fig:PEA}. 
Thus, we would expect that our approach is more attractive for implementations on devices constrained by NISQ-era resources

We note that further improvements to our method may be possible, perhaps resulting in 
the same complexity as that of the standard approach. An interesting problem is then
to understand whether all the  $\tilde g_k$'s are actually needed to solve the QEEP, or if fewer suffice. 
However, we do not expect that the largest value of $k$ needed by our method can be significantly improved
(e.g., by considering other functions $f_j(\lambda)$), since we already achieve $|k|=\tilde O(1/\epsilon)$,
and this may be optimal.

Our techniques  may also find applications in other problems, including
signal analysis. For example, given a time dependent signal $g(t)$,
we can use our analysis to obtain frequency and amplitude estimates, thereby avoiding some disadvantages arising from the use of the DFT.

%%%%%%%%%%%%%%%%%%%%%%%%%%%%%%%%%%%%%%%%%%%%%%%%%%%%%%%%%%%%%
\paragraph{\textbf{Acknowledgements.}}
We thank Yi\u{g}it Suba\c{s}\i  \ for discussions and comments.
This work was supported in part by the Laboratory Directed Research and Development program of Los Alamos National Laboratory. Los Alamos National Laboratory is managed by Triad National Security, LLC, for the National Nuclear Security Administration of the U.S.
Department of Energy under Contract No. 89233218CNA000001. Part of this work was also supported by the U.S. Department of Energy, Office of Science, Office of Advanced Scientific Computing Research, Quantum Algorithms Teams program.


%%%%%%%%%%%%%%%%%%%%%%%%%%%%%%%%%%%%%%%%%%%%%%%%%%%%%%%%
%%%%%%%%%%%%%%%%%%%%%%%%%%%%%%%%%%%%%%%%%%%%%%%%%%%%%%%%
\appendix
\section{Appendix A}
\label{app:1}

Each function $f_j(x)$ defined in Eq.~\ref{eq:fsmooth} is the convolution
of two non-negative functions and then $f_j(x) \ge 0$.
Also, since ${1}_j(x')$ is supported in $\tl_j-\epsilon/2 \le x'\le \tl_j+\epsilon/2$,
each $f_j(x)$ is identically zero if $x \le \tl_j-\epsilon$
or $x \ge \tl_j+\epsilon$, and can only be nonzero for $x \in \cV_j$.
We note that ${ 1}_j(x')+{ 1}_{j+1}(x')$ is the indicator function
supported in $\tl_j-\epsilon/2 \le x' \le \tl_{j+1}+\epsilon/2$. Since 
$\int_{-\epsilon/2}^{\epsilon/2} dx' \; h_\epsilon(x')=1$, we obtain
$f_j(x)+f_{j+1}(x)=1$ for $\tl_j \le x \le \tl_{j+1}$ or $x \in \cV_j \cap \cV_{j+1}$,
and Eq.~\ref{eq:fnormalization} is satisfied after replacing
$x$ by the eigenvalue $\lambda$.

We let $H_\epsilon(k)$ be the Fourier transform
of $h_\epsilon(x)$, $k \in \mathbb R$, so that 
\begin{equation}
    h_\epsilon(x)=\frac 1 {\sqrt{2\pi}} \int_{-\infty}^\infty dk \; H_\epsilon(k) e^{i x k} \;.
\end{equation}
Using standard properties of Fourier transforms, $H_\epsilon(k)=H(k \epsilon/2)$,
where $H(k')$ is the Fourier transform of the bump function $h(x)$.
The Fourier transform of the indicator function ${ 1}_j(x)$ is well known and
the convolution theorem implies
\begin{equation}
\label{eq:F_jFT}
  F_j(k)= 2 H(k \epsilon/2)e^{-i \tl_j k}  \frac {\sin(k \epsilon/2)}k  \;,
\end{equation}
where $F_j(k)$ is the Fourier transform of $f_j(x)$.
The Fourier series is
\begin{equation}
\label{eq:Poisson2}
    \rf_j(x) = \frac {1} {\sqrt{2 \pi}} \sum_{k}F_j(k) e^{i  x k}  \;,
\end{equation}
where the sum ranges over all integer values of $k$.
It is for this reason that we prefer the periodic function $\rf_j (x)$ rather than
$f_j(x)$; Eq.~\ref{eq:Poisson2} simplifies the following analysis.

The Fourier analysis of bump functions is 
provided in Ref.~\cite{Joh15}.
It is shown that, in the asymptotic
limit, $|H(k')|$ decays as $|k'|^{-3/4} \exp(-\sqrt{|k'|})$. For our analysis,
it suffices to assume that there exists a constant $\alpha>1$
such that, for all $|k'| \ge \alpha/2$, $|H(k')| \le
\exp(-\sqrt{|k'|})$. Then, Eq.~\ref{eq:F_jFT} implies
\begin{equation}
\label{eq:FTbound1}
    |F_j(k)| \le \epsilon \exp(-\sqrt{|k \epsilon/2|}) \;,
\end{equation}
for all $k$ that satisfy $|k| \ge \alpha/\epsilon$.
Also, note that $|H(k)| \le |H(0)|=1/{\sqrt{2\pi}}$, for all $k$,
and $|F_j(k)| \le \epsilon/(2 \pi)$ in general.

We let $ \rf '_j (x)$ be the approximate functions
that satisfy
\begin{equation}
\label{eq:approx1}
    | \rf_j (x) -  \rf '_j  (x) | \le \epsilon/(2M)   , \; \forall x \in \mathbb{R} \; , 0\le j \le M-1 \;,
\end{equation}
and $\hat \rf '_j (H)$ the corresponding operators,
obtained by replacing $x$ by $H$.
We also define the vector ${\bf p'}:=(p'_0,\ldots,p'_{M-1})$,
with 
\begin{equation}
     p'_j = \Tr [\rho . \hat \rf '_j (H)] \;.
\end{equation}
Note that,
from Eq.~\ref{eq:approx1},
\begin{align}
    |p'_j-p_j| & = \left | \Tr [\rho. (\hat \rf '_j (H)- \hat \rf _j (H))] \right| \\
    & \le  \sum_\lambda r_\lambda . |\rf '_j (\lambda) - \rf_j (\lambda)| \\
    & \le \epsilon/(2M) \; ,
\end{align}
where $r_\lambda$ is the probability of $\rho$ being in the eigenstate $\ket{\psi_\lambda}$.
Then,
\begin{equation}
    \| {\bf p'} - {\bf p} \|_1 \le \epsilon/2 \;.
\end{equation}

The functions $\rf '_j (x)$ result from approximating
$\rf_j (x)$ by a finite sum, i.e., by dropping the terms where $|k| \ge N$ in Eq.~\ref{eq:Poisson2}.
Using Eq.~\ref{eq:FTbound1} -- and therefore assuming that $N \ge \alpha/\epsilon$ -- we obtain
\begin{align}
   |\rf '_j (x) - \rf_j (x)| & \le \frac 1 {\sqrt{2 \pi}}
     \sum_{|k|\ge N} |F_j(k)| \\
     & \le (\epsilon/2)  \sum_{|k|\ge N} e^{-\sqrt{|k \epsilon/2|}} \\
     & \le 2 \int_{(N-1)\epsilon/2}^\infty dy \; e^{-\sqrt y} \\
     \label{eq:approx2}
     & \le 4 e^{-\sqrt{(N-1) \epsilon/2}}(1+\sqrt{(N-1)\epsilon/2})\;.
\end{align}
Then, there exists
$N=O(\log^2(1/\epsilon)/\epsilon)$ such that Eq.~\ref{eq:approx2}
 is bounded from above by $\epsilon/(2M)\ge \epsilon^2/4$,
and Eq.~\ref{eq:approx1} is satisfied. The constant factor hidden by the big-$O$ 
notation may be obtained from numerical simulations--see Appendix~C.

%%%%%%%%%%%%%%%%%%%%%%%%%%%%%%%%%%%%%%%%%%%%%%
\section{Appendix B}
\label{app:2}

We now prove Eq.~\ref{eq:expectederror}. First,
we note
\begin{align}
    \tau & = \sum_\lambda r_\lambda T(\lambda) \; \\
    \label{eq:taudef}
    & =  \sum_\lambda r_\lambda (\sum_{j=0}^{M-1}  f_j(\lambda) )  T(\lambda)  \\
    & = \sum_{j=0}^{M-1} \sum_{\lambda \in \cV_j} r_\lambda f_j(\lambda) T(\lambda)\;,
\end{align}
where $r_\lambda$ is the support of $\rho$ in the eigenstate $\ket{\psi_\lambda}$, and we used the properties $\sum_{j=0}^{M-1} f_j(\lambda)=1$, for all $|\lambda|\le 1/2$, and $f_j(\lambda)=0$, for $\lambda \notin \cV_j$. We also define the approximate expectation
\begin{equation}
     \tau' := \sum_{j=0}^{M-1} p_j T(\tilde \lambda_j) \;,
\end{equation}
 where
\begin{equation}
    p_j = \sum_{\lambda \in \cV_j} f_j(\lambda) r_\lambda  ; 
\end{equation}
see Eq.~\ref{eq:p_jdef}. We will use $|\tilde \tau - \tau| \le |\tilde \tau - \tau '| +|\tau ' - \tau|$ to obtain Eq.~\ref{eq:expectederror}. Then,
\begin{equation}
    \tau'- \tau = \sum_{j=0}^{M-1} \sum_{\lambda \in \cV_j}  f_j(\lambda) r_\lambda (T(\tilde \lambda_j) - T(\lambda)) \;.
\end{equation}
We define 
\begin{equation}
    z:=\sup_{|x|\le 1/2,|x'|\le \epsilon} |T(x+x')-T(x)| \;,
\end{equation}
so that
\begin{equation}
    |T(\tilde \lambda_j) - T(\lambda)| \le z \;,
\end{equation}
if $\lambda \in \cV_j$. Then,
\begin{align}
   |  \tau'- \tau | &\le z \sum_{j=0}^{M-1} \sum_{\lambda \in \cV_j}  f_j(\lambda)  r_\lambda\\
   \label{eq:tau'approx}
   & \le z \;.
\end{align}
When $T(\lambda)$ is differentiable, we can use the mean value theorem
to obtain
\begin{equation}
\label{eq:MVT}
    z \le \epsilon \max_{|\lambda|\le 1/2} |\partial_\lambda T(\lambda)| \;.
\end{equation}
The combination of Eqs.~\ref{eq:tau'approx} and~\ref{eq:MVT} yields
the second term in the right hand side of Eq.~\ref{eq:expectederror}.

The approximate expectation is 
\begin{equation}
    \tilde \tau = \sum_{j=0}^{M-1} q_j T(\tilde \lambda_j) \;,
\end{equation}
and then
\begin{align}
   | \tilde \tau- \tau' | & = |\sum_{j=0}^{M-1} (q_j-p_j) T(\tilde \lambda_j)| \\
   & \le \max_{|\lambda|\le 1/2} |T(\lambda)| \sum_{j=0}^{M-1} |q_j-p_j| \\
   \label{eq:tildetauapprox}
    & \le \max_{|\lambda|\le 1/2} |T(\lambda)| \| {\bf q} - {\bf p} \|_1 \;.
\end{align}
The combination of Eqs.~\ref{eq:tildetauapprox} and~\ref{eq:QEEPsol} yields
the first term in the right hand side of Eq.~\ref{eq:expectederror}.

%%%%%%%%%%%%%%%%%%%%%%%%%%%%%%%%%%%%%%%%%%%%%%
\section{Appendix C}
\label{app:3}

The data used for Fig.~\ref{fig:NumSim} is as follows.  We set $D=5$ and obtained the eigenvalues and probabilities randomly.
 We set the precision parameter $\epsilon=0.005$ which results in $M=1+1/\epsilon=201$. We also used $N=\lceil \log^2(M)M/10 \rceil=566$.
 While the analysis in Appendix A indicates that $O(\log^2(M)M)$ suffices, we observed from numerical simulations
 that a prefactor $1/10$ suffices for our goal precision. 
 The  noise of the signal was simulated by sampling the magnitude of each $\eta_k \in \mathbb C$
uniformly from $[0,\epsilon']$ and by sampling the phase uniformly from $[0, 2 \pi]$.
In our case, we used $\epsilon'=\epsilon$.
 Let $\Delta_{\rm TS} = (\tau - \tilde \tau_{\rm TS})/\epsilon$
and $\Delta_{\rm MP} = (\tau - \tilde \tau_{\rm MP})/\epsilon$, where $\tau=\Tr[\rho . H^s]$ and $\tilde \tau_{\rm TS}$ and $\tilde \tau_{\rm MP}$ are given in Eqs.~\ref{eq:ttTS} and~\ref{eq:ttMP}, respectively. After 5 simulation runs, the obtained results are as follows. For $s=1$, we obtained $\Delta_{\rm TS} = \{-0.683, -0.116, -0.160, 0.024, -0.222 \}$ and 
$\Delta_{\rm MP}=\{ 1.116, 1.191, -2.703, 0.355, 0.959 \}$.
For $s=2$, we obtained $\Delta_{\rm TS} = \{0.267, 0.019, 0.027, 0.036, -0.052 \}$ and 
$\Delta_{\rm MP}=\{ 1.687, -4.005, -0.483, 1.782, -0.521 \}$.
For $s=4$, we obtained $\Delta{\rm TS} = \{0.067, 0.003, 0.003, 0.010, -0.013 \}$ and 
$\Delta_{\rm MP}=\{ 19.175, -32.108, 8.724, 23.022, 6.335 \}$.
For the example shown in Fig.~\ref{fig:NumSimQEEP}, the values of $(\lambda,r_\lambda)$
are (-0.134, 0.33), (-0.130, 0.08), (0.208, 0.20), (0.408, 0.18), and (0.438, 0.21).


\begin{thebibliography}{39}
\bibitem [{not({\natexlab{a}})}]{note1}%
  \BibitemOpen
  \href@noop {} {\bibfield  {journal} {\bibinfo  {journal} {If $1/\epsilon$ is
  not an integer, we can define $M=1+\lceil1/\epsilon\rceil$ and assign
  $1/(M-1) \rightarrow \epsilon$.}\ } ({\natexlab{a}})}\BibitemShut {NoStop}%
\bibitem [{not({\natexlab{b}})}]{note2}%
  \BibitemOpen
  \href@noop {} {\bibfield  {journal} {\bibinfo  {journal} {The $F^j(k)$'s may
  be obtained from approximate integrations resulting in an additional
  pre-processing cost. However, this step has to be done only once as the
  functions are independent of $H$.}\ } ({\natexlab{b}})}\BibitemShut {NoStop}%
\bibitem [{not({\natexlab{c}})}]{note3}%
  \BibitemOpen
  \href@noop {} {\bibfield  {journal} {\bibinfo  {journal} {The analysis in
  Appendix~B can be applied to the case where the function $T(x)$ is not
  required to be differentiable.}\ } ({\natexlab{c}})}\BibitemShut {NoStop}%
\end{thebibliography}
\end{document}